\documentclass[a4paper, 10 pt, conference]{ieeeconf}  % Comment this line out if you need a4paper

\IEEEoverridecommandlockouts                              % This command is only needed if 
\usepackage{graphics} % for pdf, bitmapped graphics files
\usepackage{amsmath} % assumes amsmath package installed
\usepackage{amssymb}  % assumes amsmath package installed
\usepackage{cite}
\usepackage{graphicx}
\usepackage{booktabs}
\usepackage{multirow}
\usepackage{threeparttable}

\title{\LARGE \bf
VAE-Based Synthetic EMG Generation with Mix-Consistency Loss for Recognizing Unseen Motion Combinations
}

\author{Itsuki Yazawa$^{1}$ and Akira Furui$^{1}$% <-this % stops a space
\thanks{This work was partially supported by JSPS KAKENHI Grant Number JP23K28128.}% <-this % stops a space
\thanks{$^{1}$I. Yazawa and A. Furui are with Graduate School of Advanced Science and Engineering,
        Hiroshima University, Higashi-hiroshima, Japan 
        (e-mail: {\tt\small \{itsukiyazawa, akirafurui\}@hiroshima-u.ac.jp})}%
}

\begin{document}

\maketitle
\thispagestyle{empty}
\pagestyle{empty}

%%%%%%%%%%%%%%%%%%%%%%%%%%%%%%%%%%%%%%%%%%%%%%%%%%%%%%%%%%%%%%%%%%%%%%%%%%%%%%%%
\begin{abstract}

Electromyogram (EMG)-based motion classification using machine learning has been widely employed in applications such as prosthesis control.
While previous studies have explored generating synthetic patterns of combined motions to reduce training data requirements, these methods assume that combined motions can be represented as linear combinations of basic motions.
However, this assumption often fails due to complex neuromuscular phenomena such as muscle co-contraction, resulting in low-fidelity synthetic signals and degraded classification performance.
To address this limitation, we propose a novel method that learns to synthesize combined motion patterns in a structured latent space. 
Specifically, we employ a variational autoencoder (VAE) to encode EMG signals into a low-dimensional representation and introduce a mix-consistency loss that structures the latent space such that combined motions are embedded between their constituent basic motions. 
Synthetic patterns are then generated within this structured latent space and used to train classifiers for recognizing unseen combined motions.
We validated our approach through upper-limb motion classification experiments with eight healthy participants. The results demonstrate that our method outperforms input-space synthesis approaches, achieving approximately $30\%$ improvement in accuracy.

\end{abstract}

%%%%%%%%%%%%%%%%%%%%%%%%%%%%%%%%%%%%%%%%%%%%%%%%%%%%%%%%%%%%%%%%%%%%%%%%%%%%%%%%
\section{Introduction}

Electromyogram (EMG) signals are electrical signals generated by muscle activity, measured non-invasively using surface electrodes. 
These signals contain rich information about human motor intentions and are commonly used for control interfaces, particularly prosthetic limbs~\cite{farina_extraction_2014}.
Recent advances in deep learning have significantly improved the accuracy of EMG-based motion classification~\cite{cote-allard_deep_2019,triwiyanto2020improved}.

However, EMG-based motion classification faces a fundamental challenge: the training data requirement scales with the number of target motions.
This challenge becomes particularly acute for combined motions---simultaneous executions of multiple basic motions---which increase combinatorially. For instance, while four basic upper-limb motions require four data collection sessions, their pairwise combinations require an additional six sessions.
This combinatorial growth in data collection leads to user fatigue and limits the practical deployment of multi-degree-of-freedom (DoF) control systems~\cite{nagata_basic_2011,olsson_extraction_2019}.

To address this challenge, researchers have proposed representing combined motions as combinations of basic motion patterns, enabling multi-DoF control with limited training data~\cite{shima_classification_2010}.
To improve the robustness of such approaches, recent methods generate synthetic combined motion data from basic motion measurements~\cite{olsson_can_2021, yazawa_recognition_2025}.
Our previous work~\cite{yazawa_recognition_2025} demonstrated that convex combinations of basic motion patterns in the input signal space could improve classification accuracy for unseen combined motions.
However, this approach relies on the assumption that combined motion patterns can be represented as linear combinations of basic motion patterns---an assumption that often fails to capture complex neuromuscular phenomena such as muscle co-contraction and motion-dependent modulation patterns.

% Among these, one promising approach is to generate synthetic data for combined motions by mixing the EMG signals of basic motions, and using this data to train classifier, thereby improving performance on previously unseen combined motions~\cite{olsson_can_2021, yazawa_recognition_2025}.
% For example, Olson \textit{et al.} proposed a method that uses a Variational Autoencoder (VAE)~\cite{kingma_auto-encoding_2013} to obtain latent representations of basic motions and linearly mix them to generate synthetic data for unseen combined motions.

% Nevertheless, these methods generally assume that combined motions can be represented as linear combinations of basic motions. 
% This assumption often fails in the presence of nonlinear relationships, such as co-contraction patterns, resulting in poor-quality synthetic data and degraded classification performance.
% To enable more flexible and accurate synthesis, it is necessary to learn the combination mechanism from real basic and combined motion data.
% Once acquired, this mechanism can be applied to generate synthetic data under new environments or conditions.

In this paper, we propose a novel approach that learns to synthesize combined motion patterns in a structured latent space. 
We employ a variational autoencoder (VAE)~\cite{kingma_auto-encoding_2013} to encode EMG signals into a low-dimensional representation and introduce a mix-consistency loss that structures the  latent space such that combined motions are embedded between their constituent basic motions.
Synthetic combined motion data generated from this space are then used together with measured basic motion data to train a classifier, thereby enabling recognition of unseen combined motions without requiring their explicit measurement.

% The ultimate goal of this work is to establish an inter-subject classification framework that generalizes to new users. 
% By training the generative model on data from a small number of subjects, we aim to classify unseen combined motions for new users using only basic motion data. 
% As a first step toward this goal, this paper focuses on structuring the latent space using the VAE and the mix-consistency loss, and generating synthetic data within that space. 
% To verify the effectiveness of the proposed method, we evaluate its classification performance and data generation capability through intra-subject experiments.
% いらないかも
% As a step toward inter-subject generalization, our approach is designed to learn a reusable motion-combination mechanism from a small number of subjects. 
% This allows classification of unseen combined motions for new users based only on their basic motion data. 
% We validate the proposed method through intra-subject experiments, evaluating both classification performance and data generation quality.

Our long-term vision is to develop a generalizable synthesis model that can be pre-trained on multiple subjects and transferred to new users, ultimately eliminating the need for combined motion data collection from individual users. 
As a first step toward this goal, we validate the feasibility of latent-space synthesis through intra-subject experiments. 

\section{Proposed Method}

We address the problem of classifying EMG patterns $\mathbf{X}\in\mathbb{R}^{W\times D}$ measured from $D$ electrodes with window size $W$.
The target motion classes consist of $M_b$ basic motions and $M_c$ combined motions, where each combined motion represents a combination of $K_m$ basic motion classes.

Our approach consists of two sequential training phases.
In the VAE training phase, we collect EMG signals for all motion classes, including both basic and combined motions, to train a VAE capable of synthesizing user-specific EMG patterns for combined motions.
Subsequently, in the classifier training phase, we measure EMG signals only for basic motion classes and input these into the pre-trained VAE to synthesize EMG patterns for the combined motions. 
The classifier is then trained using real EMG patterns for basic motions and synthetic EMG patterns for combined motions.

\subsection{Training of VAE for Synthetic EMG Patterns}

\begin{figure*}[t]
		\centering
		\includegraphics[keepaspectratio, width=0.85\linewidth]{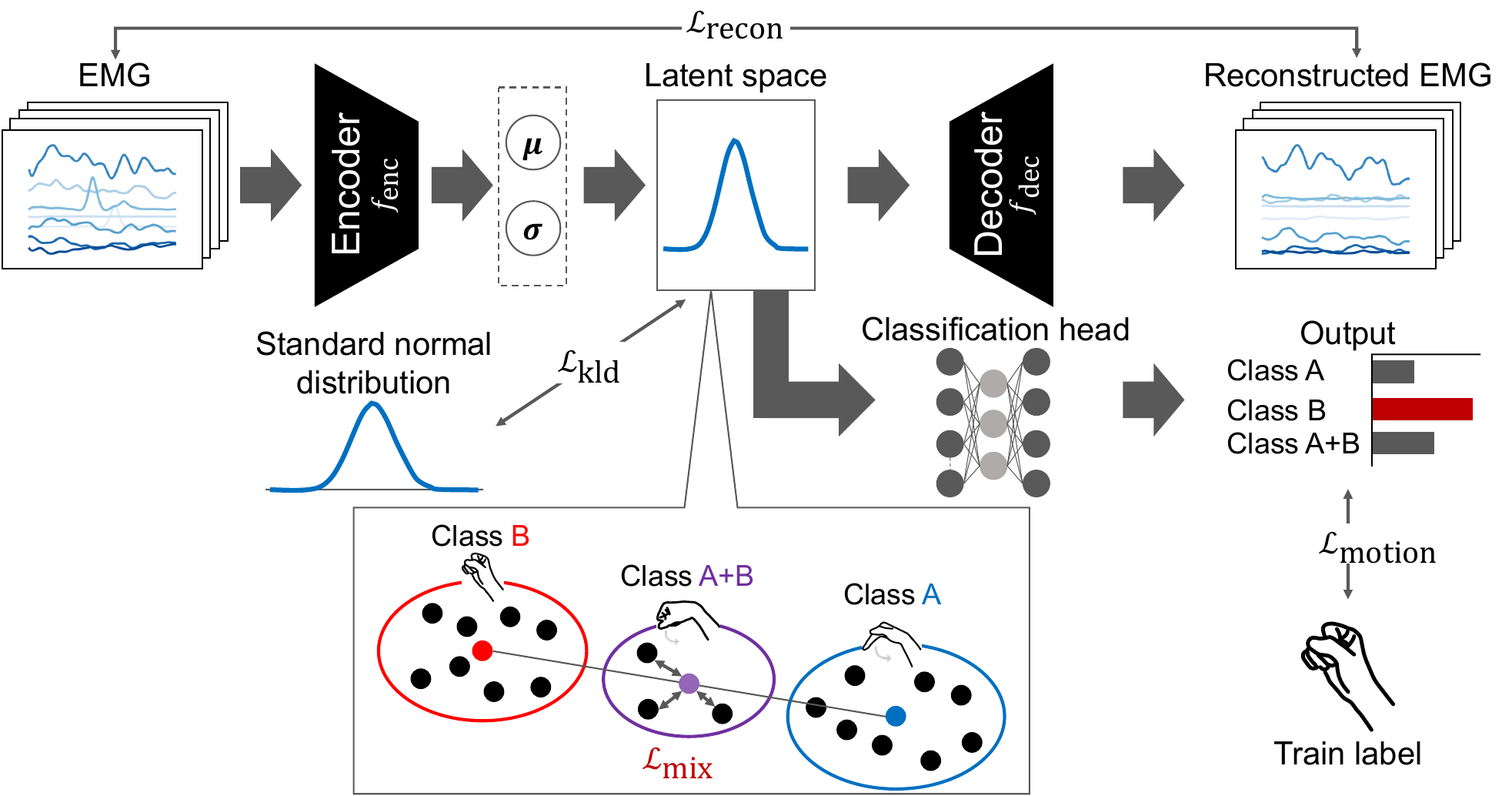}
	\caption{Overview of the proposed VAE-based framework for synthetic EMG generation.
    The mix-consistency loss $\mathcal{L}_\mathrm{mix}$ structures the latent space so that combined motions are positioned between their constituent basic motions, enabling realistic synthetic pattern generation.}
	\label{overview}
\end{figure*}

Fig.~\ref{overview} illustrates an overview of the proposed method.
We first train a VAE using dataset $\mathcal{D}_{1}=\{(\mathbf{X}_i, \mathbf{y}_i)\}^{N_{1}}_{i=1}$, which contains EMG signals for both basic and combined motions. 
Here, $\mathbf{X}_i\in\mathbb{R}^{W\times D}$ represents the EMG pattern for the $i$-th sample, and $\mathbf{y}_i\in\{0,1\}^{M_b+M_c}$ is a one-hot encoded label vector. 
The VAE encodes each EMG signal $\mathbf{X}_i$ into a $Z$-dimensional latent representation $\mathbf{z}_i\in\mathbb{R}^Z$.

The proposed VAE comprises three modules:
\begin{itemize}
    \item \textbf{Encoder} extracts the mean $\boldsymbol{\mu}_i \in \mathbb{R}^Z$ and standard deviation $\boldsymbol{\sigma}_i \in \mathbb{R}^Z$ of the approximate posterior distribution $q(\mathbf{z}_i|\mathbf{X}_i)$ from input EMG signal $\mathbf{X}_i$.
    It consists of 1D convolutional layers followed by fully connected layers, incorporating batch normalization, max-pooling, and Leaky ReLU activations.
    
    \item \textbf{Decoder} samples a latent representation $\mathbf{z}_i \sim \mathcal{N}(\boldsymbol{\mu}_i, \text{diag}(\boldsymbol{\sigma}_i^2))$ and reconstructs EMG signal $\hat{\mathbf{X}}_i\in\mathbb{R}^{W\times D}$ from the latent vector $\mathbf{z}_i$.
    It comprises a fully connected layer followed by 1D transposed convolutional layers with progressive upsampling and nonlinear transformations using Leaky ReLU activations and a final hyperbolic tangent function.
    
    \item \textbf{Classification head} predicts class label $\hat{\mathbf{y}}_i\in\{0,1\}^{M_b+M_c}$ from latent representation $\mathbf{z}_i$ using a multilayer perceptron.
\end{itemize}

To train the VAE and structure the latent space, we minimize the following total loss function:
\begin{align}
    \mathcal{L}_\text{vae} = \mathcal{L}_{\text{recon}} + \beta \mathcal{L}_{\text{kld}} + \gamma \mathcal{L}_{\text{motion}} + \mathcal{L}_{\text{mix}},
\end{align}
where $\beta$ and $\gamma$ are hyperparameters controlling the respective loss terms.

The total loss consists of four components:
\begin{itemize}
    \item \textbf{Reconstruction loss} encourages accurate reconstruction of input EMG signals $\mathbf{X}_i$ from the reconstructed outputs $\hat{\mathbf{X}}_i$.
    This loss comprises time-domain mean squared error (MSE) and frequency-domain MSE based on the fast Fourier transform (FFT):
    \begin{align}
        \mathcal{L}_{\text{recon}} &= \mathcal{L}_{\text{time}} + \delta \mathcal{L}_{\text{freq}}, \\
        \mathcal{L}_{\text{time}} &= \frac{1}{N_{1}DW}\sum^{N_{1}}_{i=1}\left\|\hat{\mathbf{X}}_{i}-\mathbf{X}_{i}\right\|^2_{2},\\
        \mathcal{L}_{\text{freq}} &= \frac{1}{N_{1}D}\sum_{i=1}^{N_{1}}\sum_{d=1}^{D}\left\|\text{FFT}(\mathbf{x}_{i,d})-\text{FFT}(\hat{\mathbf{x}}_{i,d})\right\|^2_2.
    \end{align}
    Here, $\mathbf{x}_{i,d} \in \mathbb{R}^W$ represents the one-dimensional time series corresponding to the $d$-th channel of the $i$-th sample, where $\mathbf{X}_i= [\mathbf{x}_{i, 1}, \mathbf{x}_{i, 2}, \ldots, \mathbf{x}_{i, D}]^\top$.
    The FFT is computed for each channel over the entire time window, with weighting parameter $\delta$ controlling the contribution of the frequency-domain loss.

    \item \textbf{Kullback-Leibler divergence} encourages the posterior distribution to match a standard normal distribution:
    \begin{align}
        \mathcal{L}_{\text{kld}} = \frac{1}{2N_{1}} \sum^{N_{1}}_{i=1}\sum^{Z}_{j=1} (\mu_{ij}^2+\sigma^2_{ij}-\log\sigma^2_{ij}-1).
    \end{align}

    \item \textbf{Motion classification loss} promotes class-discriminative structure in the latent space:
    \begin{align}
        \mathcal{L}_{\text{motion}} = \frac{1}{N_1}\sum_{i=1}^{N_1}\ell(\hat{\mathbf{y}}_i, \mathbf{y}_i),
    \end{align}
    where $\hat{\mathbf{y}}_i$ is the predicted class probability from the classification head given latent vector $\mathbf{z}_i$, and $\ell(\cdot,\cdot)$ denotes the cross-entropy loss.

    \item \textbf{Mix-consistency loss} encourages combined motions to be embedded near the centroid of their constituent basic motions:
    \begin{align}
        \mathcal{L}_{\text{mix}} = \frac{1}{|M_c|}\sum_{c \in M_c}\frac{1}{|I_c|}\sum_{i\in I_c}\left\|\mathbf{z}_{i}^{(c)}-\frac{1}{K_m}\sum_{b=1}^{K_m}\bar{\mathbf{z}}^{(b)}\right\|^2_2.
    \end{align}
    Here, $I_c$ denotes the set of sample indices for combined motion class $c$, $\mathbf{z}_{i}^{(c)}$ is the latent vector of the $i$-th sample in class $c$, and $\bar{\mathbf{z}}^{(b)}$ is the mean latent vector of basic motion class $b$ computed over all samples in $\mathcal{D}_1$.
\end{itemize}

\subsection{Generation of Synthetic EMG and Classifier Training}

Using the pre-trained VAE, we perform data generation and classifier training based on a newly measured dataset $\mathcal{D}_{2} = \{(\mathbf{X}_i, \mathbf{y}_i)\}^{N_{2}}_{i=1}$, which contains only basic motion classes. 
Each label $\mathbf{y}_i\in\{0,1\}^{M_b}$ is a one-hot vector corresponding to a basic motion.

\begin{figure}[t]
		\centering
		\includegraphics[width=\linewidth]{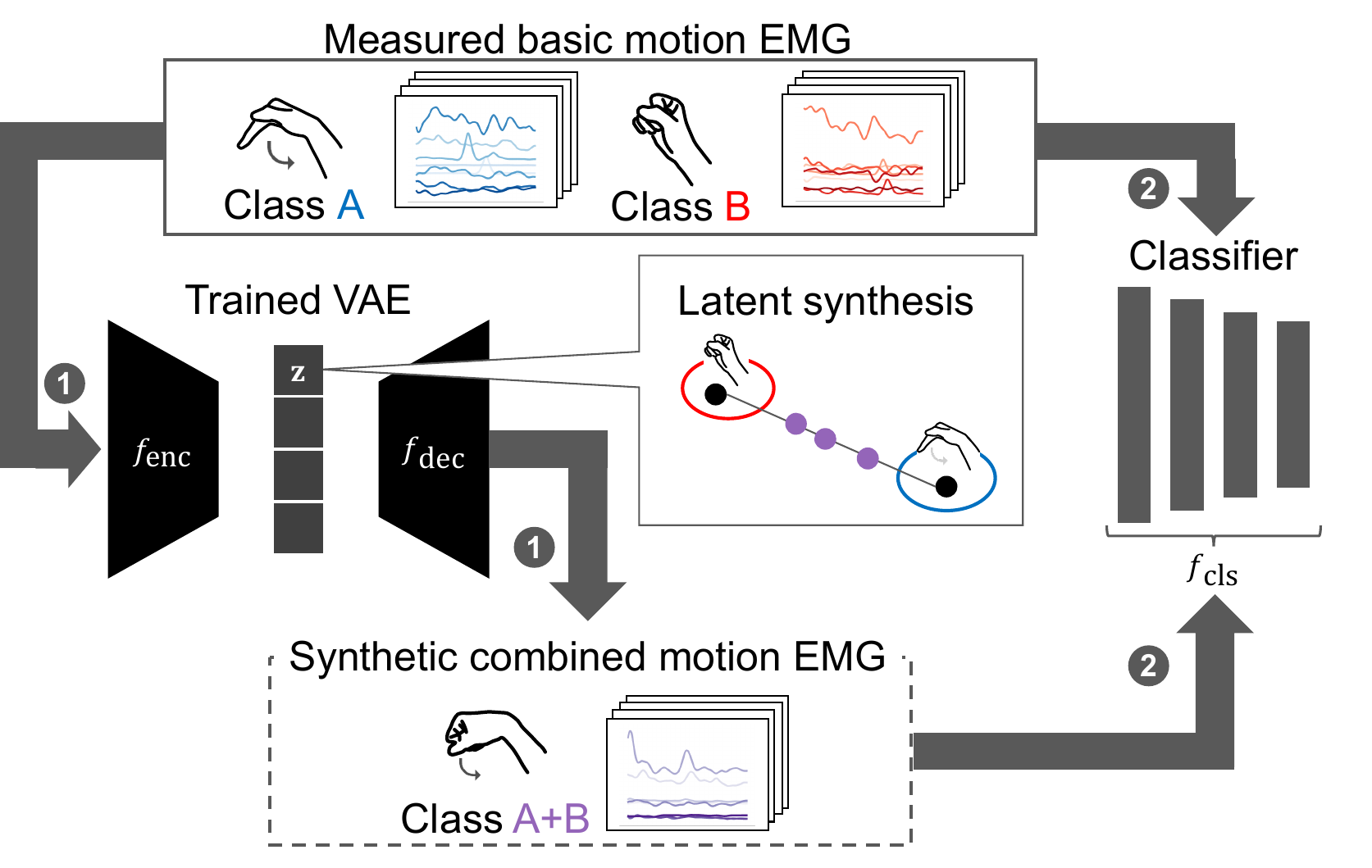}
	\caption{Pipeline for synthetic combined motion generation and classifier training using measured basic motion EMG and pre-trained VAE latent synthesis.}
	\label{generation}
\end{figure}

To generate combined motion samples, we employ Mixup-like convex combination in the latent space of the VAE.
Fig.~\ref{generation} illustrates the synthetic data generation method.
We randomly sample $K_m$ different basic motion samples and linearly combine their latent representations to generate synthetic data for combined motion class $m$:
\begin{align}
    \boldsymbol{\mu}_i^{(k)}, \boldsymbol{\sigma}_i^{(k)} &= f_\text{enc}(\mathbf{X}_{i}^{(k)}),\\
    \mathbf{z}_{i}^{(k)} &\sim \mathcal{N}\left(\boldsymbol{\mu}_i^{(k)}, \text{diag}({\boldsymbol{\sigma}_i^{(k)}}^2)\right), \\
	\tilde{\mathbf{z}}_{i}^{(m)} &= \sum^{K_m}_{k=1}\lambda_k\mathbf{z}_{i}^{(k)}, \\
	\tilde{\mathbf{X}}_{i}^{(m)} &= f_\text{dec}(\tilde{\mathbf{z}}_{i}^{(m)}).
\end{align}
Here, $\mathbf{X}_{i}^{(k)}$ is the EMG signal for the $i$-th sample from basic motion class $k$, and $f_{\text{enc}}(\cdot), f_{\text{dec}}(\cdot)$ denote the encoder and decoder of the VAE, respectively.
The coefficients $\lambda_k\in[0,1]$ represent mixing ratios that satisfy $\sum_{k=1}^{K_m}\lambda_k=1$.
These coefficients are sampled from a symmetric Dirichlet distribution to incorporate randomness while preserving the convex combination constraint:
\begin{align}
    p(\boldsymbol{\lambda} \mid \alpha) &= \frac{1}{B(\mathbf{\alpha})} \prod_{k=1}^{K_m} \mathbf{\lambda}_{k}^{\alpha-1}, \\
    B(\mathbf{\alpha}) &= \frac{\Gamma(\alpha)^{K_m}}{\Gamma(K_m \alpha)},
\end{align}
where $\boldsymbol{\lambda}=[\lambda_1, \lambda_2, \ldots, \lambda_{K_m}]^\top$ and $\alpha>0$ controls the shape of the Dirichlet distribution. 
The label $\hat{\mathbf{y}}\in\{0,1\}^{M_c}$ for each generated sample $\tilde{\mathbf{X}}_i$ is set as a one-hot vector corresponding to the target combined motion class.

After generating the synthetic data, we combine the measured basic motion dataset $\mathcal{D}_{2}$  with the synthetic combined motion dataset $\tilde{\mathcal{D}}_{2}=\{(\tilde{\mathbf{X}}_j, \tilde{\mathbf{y}}_j)\}^{\tilde{N}_2}_{j=1}$ to form the full training set $\mathcal{D}=\mathcal{D}_{2}\cup \tilde{\mathcal{D}}_{2}$.
The classifier is trained by minimizing the following loss:
\begin{align}
    \mathcal{L}_{\text{cls}} = \frac{1}{N_2} \sum_{i=1}^{N_2} \ell(f_{\text{cls}}(\mathbf{X}_i), \mathbf{y}_i) + \frac{1}{\tilde{N}_2} \sum_{j=1}^{\tilde{N}_2} \ell(f_{\text{cls}}(\tilde{\mathbf{X}}_j), \tilde{\mathbf{y}}_j),
\end{align}
where $f_{\text{cls}}(\cdot)$ represents the classifier, $\tilde{N}_2$ is the number of synthetic combined motion samples.

In summary, our approach first learns a structured latent space by training a VAE on both basic and combined motion classes. 
Combined motion samples are then generated in the latent space by convexly combining the latent vectors of $K_m$ basic motions.
Finally, a classifier is trained using both measured basic motions and synthetic combined motions, enabling recognition of previously unseen combined motions.

\section{Experiments}

\subsection{Experimental Setup} 
To evaluate the effectiveness of the proposed method, we conducted an upper-limb motion classification experiment using an EMG dataset from our previous work~\cite{yazawa_recognition_2025}.
The dataset includes recordings from eight healthy adult participants (mean age: $23\pm0.76$ years).
For each participant, eight surface electrodes were placed circumferentially around the right forearm near the elbow ($D=8$).
EMG signals were acquired using a wireless acquisition system (Trigno, Delsys Inc.) at a sampling rate of 2,000 Hz.

During the experiment, participants remained seated with their right elbow resting on a desk and performed a total of 18 distinct upper-limb motions.
These included 6 basic motions:
hand opening (S1), hand grasping (S2), wrist extension (S3), wrist flexion (S4), pronation (S5), and supination (S6) and 12 combined motions, each constructed by combining two basic motions ($K_m=2$): opening and pronation (C1), grasping and pronation (C2), extension and pronation (C3), flexion and pronation (C4), opening and supination (C5), grasping and supination (C6), extension and supination (C7), flexion and supination (C8), opening and extension (C9), grasping and extension (C10), opening and flexion (C11), and grasping and flexion (C12).
Thus, the total number of classes was $M_b=6$ and $M_c=12$.

Each motion was performed for 4 seconds continuously, and the task was repeated across 6 trials. 
Participants rested for 4 seconds between motions and 40 seconds between trials while maintaining a relaxed posture.
Prior to the study, participants provided written consent after being briefed about the research objectives. 
The study protocol received approval from the Ethics Committee at Hiroshima University (approval number: E-840).

\subsection{Data Processing}
The raw EMG signals were preprocessed as follows.
First, a fourth-order band-stop filter ($60$--$62$ Hz) was applied to eliminate power line noise. 
Subsequently, full-wave rectification was performed, followed by smoothing using a second-order Butterworth low-pass filter with a cutoff frequency of $2.0$ Hz.
To remove transitional artifacts at motion onset, the initial $5\%$ of each signal segment was discarded. 
The smoothed signals were then clipped between the $1$st and $99$th percentiles to suppress outliers, then normalized to the range $[-1, 1]$.
The resulting EMG signals were used as the input patterns $\mathbf{X}$.

The experimental protocol employed a cross-validation approach using the $6$ available trials per participant.
For each fold, trials were allocated as follows:
\begin{itemize}
    \item Two trials containing both basic and combined motions were used to train the VAE.
    \item Two different trials containing only basic motions were used to generate synthetic combined motion data and train the classifier.
    \item The remaining $2$ trials containing both basic and combined motions served as test data.
\end{itemize}
This allocation ensures that the classifier never sees actual combined motion data during training, thereby validating the method's ability to recognize unseen combined motions.
Classification accuracy was computed by averaging results across all possible trial combinations.

\subsection{Evaluation Methods}

\subsubsection{Training configuration}
We used a convolutional neural network (CNN) as the classifier.
Both the VAE and CNN classifier were optimized using Adam with a batch size of $64$ and a learning rate of $1.0 \times 10^{-4}$.
Training proceeded for 200 epochs for the VAE and 10 epochs for the CNN classifier.
Detailed model architectures are provided in Table.~\ref{tab:model}.
In our model, the parameters $\alpha$ of all Leaky ReLU functions were set to $0.2$.
Moreover, the size of latent vector was set to 6 ($Z=6$).

The loss function hyperparameters were configured as follows.
The $\beta$ parameter was scaled by the latent-to-input dimensionality ratio to maintain consistent loss scales across different dimensions~\cite{higgins_beta-vae_2017}:
\begin{align}
    \beta = \beta_0 \cdot \frac{Z}{WD},
\end{align}
where $\beta_0$ represents the base value that was gradually increased during training to stabilize VAE optimization, with a maximum value of $0.3$.
The $\gamma$ parameter was fixed at 2 throughout training, while $\delta$ was set to 0 for the first 100 epochs and changed to 1 for the remaining training.
For synthetic data generation, the $\alpha$ parameter of the Dirichlet distribution was set to $50$.

\subsubsection{Baseline comparisons}

We evaluated the proposed method against three baseline approaches:
\begin{itemize}
    \item \textbf{Fully-supervised:}
    Trained on measured data from $2$ trials of all $18$ motion classes (basic and combined).
    This represents the ideal scenario with complete supervision. 

    \item \textbf{Basic-only~\cite{shima_classification_2010}:}
    Trained exclusively on measured data from $2$ trials of the 6 basic motion classes, providing a baseline for methods without combined motion data.
    
    \item\textbf{Mixup-based synthesis~\cite{yazawa_recognition_2025}:}
    Applied data augmentation method by mixing input signals directly based Mixup concept~\cite{zhang_mixup_2017} without using the VAE, as proposed in our previous work. This represents a simpler synthesis approach for comparison.
\end{itemize}
All baseline methods used identical training configurations to ensure fair comparison.

\begin{table*}[t]
	\caption{Model architecture configurations}
	\label{tab:model}
	\centering
    \begin{threeparttable}
    \begin{tabular}{llllll}
		\toprule
		Component & Layer Configuration & Output Channels & Kernel/Stride & Activation & Additional Features \\
		\midrule
		\multirow{9}{*}{VAE} & \textit{Encoder:} & & & & \\
		 & \quad 3 $\times$ 1D Conv & 32, 64, 128 & 3/1, 3/1, 4/2 & Leaky ReLU & BatchNorm; MaxPool\tnote{a} \\
		 & \quad Fully connected & 128 & -- & Leaky ReLU & -- \\
		 & \qquad $\to$ $\mu$ branch & 6 & -- & -- & -- \\
		 & \qquad $\to$ $\log \sigma^2$ branch & 6 & -- & -- & -- \\
		 & \textit{Decoder:} & & & & \\
		 & \quad Fully connected & 768 & -- & Leaky ReLU & -- \\
		 & \quad 3 $\times$ 1D TransConv & 64, 32, 8 & 4/2, 3/2, 2/2 & Leaky ReLU; Tanh\tnote{c} & BatchNorm\tnote{a}\\
		 & \textit{Classification head:} & & & & \\
		 & \quad 3 $\times$ Fully connected & 64, 32, 18 & -- & ReLU & -- \\
		\cmidrule{1-6}
		\multirow{2}{*}{CNN Classifier} & 4 $\times$ 1D Conv & 16, 32, 64, 128 & 3/1 & PReLU & BatchNorm; MaxPool; GAP \\
		 & 2 $\times$ Fully connected & 64, 18 & -- & -- & BatchNorm\tnote{b} \\
		\bottomrule
	\end{tabular}
	\begin{tablenotes}
		\item ${}^{\text{a}}$Applied to first two layers only.
		 ${}^{\text{b}}$Applied to first FC layer only.  
		 ${}^{\text{c}}$Leaky ReLU for first two layers, Tanh for final layer.
		 The abbreviations Conv, TransConv, BatchNorm, MaxPool, and GAP denote convolutional, transposed convolutional, batch normalization, max pooling, global average pooling, respectively.
	\end{tablenotes}
    \end{threeparttable}
\end{table*}

\section{Results}
\subsection{Evaluation of Synthetic EMG Pattern Generation}
\begin{figure}[t]
		\centering
		\includegraphics[width=\linewidth]{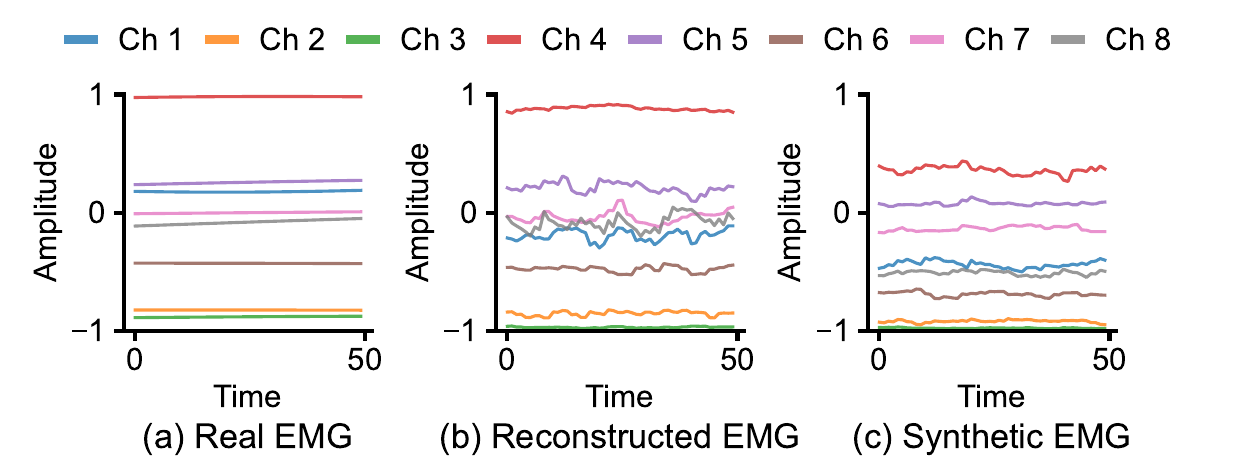}
	\caption{Example of EMG signals for combined motion C2, participant 1: (a) measured, (b) VAE-reconstructed, and (c) synthetic patterns showing comparable amplitude trends across 8 channels.}
	\label{signal}
\end{figure}

Fig.~\ref{signal} presents a comparison of EMG signals for a representative combined motion class, showing (a) the original measured EMG signal, (b) its VAE reconstructed, and (c) the reconstruction generated from synthetic data.
The horizontal and vertical axes represent time and signal amplitude, respectively. 

The comparison between Fig.~\ref{signal}(a) and (b) demonstrates that the VAE successfully captures the overall amplitude trends of the original signal.
However, minor reconstruction errors are evident in localized amplitude fluctuations and noise characteristics.
The comparison between Fig.~\ref{signal}(b) and (c) reveals that synthetic combined data produce reconstructions similar to those of actual measured data, indicating that the VAE effectively captures realistic and class-consistent features of combined motions.

\subsection{Structure of Latent Space}
\begin{figure}[t]
		\centering
		\includegraphics[keepaspectratio, width=0.9\linewidth]{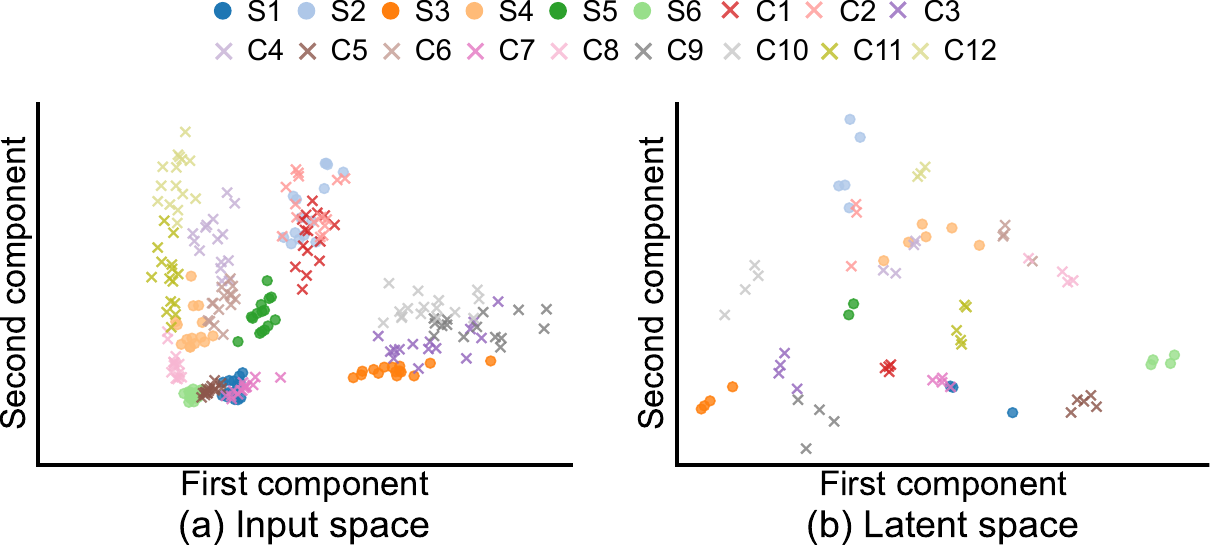}
	\caption{Principal component analysis of EMG patterns in (a) input space and (b) VAE latent space, showing structured embedding of combined motions between basic motions (Participant 1).}
	\label{real_latent}
\end{figure}

Fig.~\ref{real_latent} shows a comparison between the measured input EMG signals and their corresponding latent representations obtained from the VAE, visualized in two dimensions using principal component analysis.

% いらないかも
While all motion classes are included in Fig.~\ref{real_latent}, we highlight the combined class ``Hand Opening + Pronation'' (C1), which is composed of two basic motions: ``Opening'' (S1) and ``Pronation'' (S5).
% ここまで
In Fig.~\ref{real_latent}(a), the input EMG signals corresponding to the combined motion form a distribution that is clearly distinct from those of the individual basic motions.
This suggests that, in the original signal space, the combined motion exhibits a unique pattern not easily interpreted as a simple combination of the constituent basic motions.
In contrast, Fig.~\ref{real_latent}(b) shows the latent space, where the combined motion is distributed near the midpoint between the two corresponding basic motion clusters. 
This indicates that the latent space has been structured such that combined motions are positioned in a meaningful location—approximately interpolated between the latent representations of the basic motions involved.

\subsection{Classification Performance}
Table~\ref{table_accuracy} presents the classification accuracy results for all evaluated method.
The fully-supervised method, trained on actual EMG patterns for all motion classes, achieved a high average classification accuracy exceeding $80\%$. 
This represents an upper-bound performance under ideal conditions where complete training data are available.
In contrast, the Mixup method, which generates synthetic combined motion data in the input space, achieved lower accuracy of approximately $40\%$.
Our proposed method, utilizing a structured latent space for synthetic data generation, substantially improved classification accuracy to around $75\%$, demonstrating clear superiority over the input-space synthesis approach.

\begin{figure}[t]
		\centering
		\includegraphics[keepaspectratio, width=0.8\linewidth]{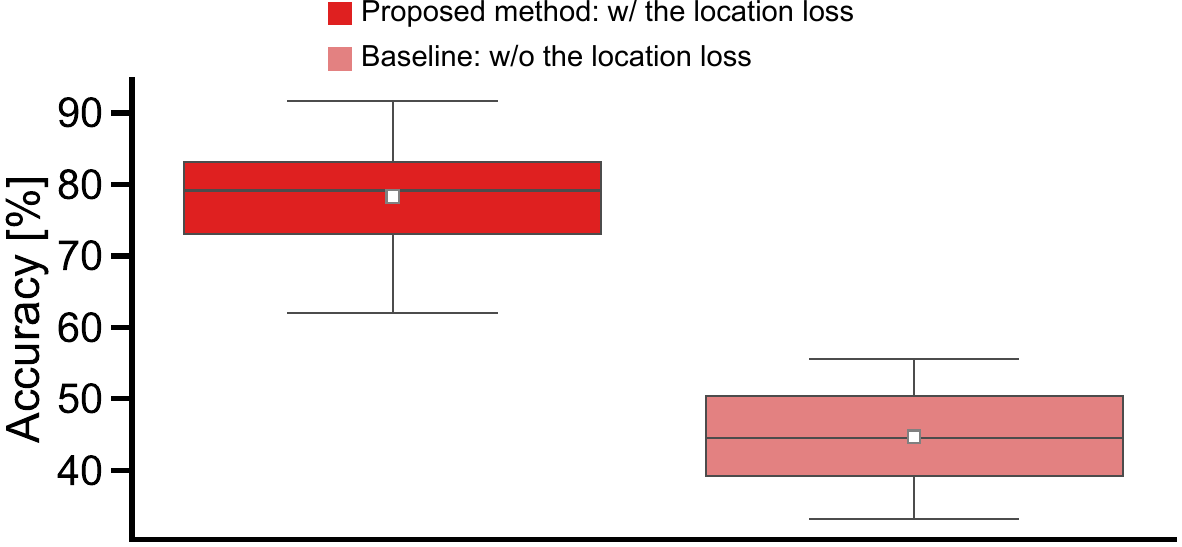}
	\caption{Ablation study of mix-consistency loss on classification accuracy.}
	\label{accuracy_ablation}
\end{figure}

Fig.~\ref{accuracy_ablation} presents an ablation study evaluating the contribution of the mix-consistency loss $\mathcal{L}_{\text{mix}}$.
The medial line and white square denote the median and mean values across participants, respectively.
The baseline model without this loss component achieved approximately $40\%$ accuracy, similar to the Mixup method.
Incorporating the mix-consistency loss resulted in notable accuracy improvement, confirming the effectiveness of structuring the latent space for combined motion synthesis.

\begin{table*}[t]
\caption{Classification accuracy (mean $\pm$ standard deviation) for proposed method and baselines}
\label{table_accuracy}
\centering
\begin{threeparttable}
\begin{tabular}{llcccc}
\toprule
\multirow{3}{*}{Method} & \multicolumn{2}{c}{Training data configuration} & \multicolumn{3}{c}{Accuracy (\%)} \\ 
\cmidrule(lr){2-3} \cmidrule(lr){4-6}
                        & Basic & Combined & Basic & Combined & Overall \\ 
\midrule
Fully-supervised        & Real              & Real                      & 83.15 $\pm$ 6.91 & 86.47 $\pm$ 5.44 & 85.36 $\pm$ 5.10 \\
Basic-only              & Real              & ---                       & \textbf{90.65 $\pm$ 7.64} &  16.42 $\pm$ 3.52               & 41.17 $\pm$ 4.34 \\
Mixup-based synthesis~\cite{yazawa_recognition_2025}                   & Real              & Synthetic (input space)   & 86.97 $\pm$ 8.94 & 23.21 $\pm$ 5.85 & 44.46 $\pm$ 6.38 \\
\textbf{Ours}          & Real              & Synthetic (latent space)  & 82.43 $\pm$ 10.24 & \textbf{76.29 $\pm$ 9.40} & \textbf{78.34 $\pm$ 9.14} \\ 
\bottomrule
\end{tabular}
	\begin{tablenotes}
		\item \textbf{Bold} values indicate the best performance among methods that do not use real combined motion data for training.
	\end{tablenotes}
\end{threeparttable}
\end{table*}

\section{Discussion}
This study aimed to enable multi-DoF motion classification without explicit training data for combined motions by learning to generate such data synthetically in advance.
The experimental results demonstrate the effectiveness of our VAE-based approach over conventional linear synthesis methods.

The conventional Mixup method, which linearly combines basic motion data in the input space, resulted in significant degradation in classification accuracy (Table~\ref{table_accuracy}).
This performance gap can be attributed to the fundamental limitation that EMG patterns of combined motions do not necessarily exhibit linear relationships with their constituent basic motions. 
The complex neuromuscular interactions, including co-contraction patterns and synergistic activations, create nonlinear dependencies that cannot be captured through simple linear interpolation in the input space.
% Hence, linear synthesis in the input space fails to capture the complex relationships inherent in real combined motions.
% In contrast, our proposed method synthesizes combined motion data in the latent space of a VAE, achieved a substantial improvement in classification accuracy. 
% As shown in Fig.~\ref{signal}, the generated synthetic EMG patterns resemble actual patterns, enabling the classifier to effectively learn the characteristics of combined motions.

Our proposed method addresses this limitation by performing synthesis in the structured latent space of a VAE. 
The substantial improvement in classification accuracy (approximately 30\% over the Mixup method) demonstrates that the learned latent representation better captures the underlying motion patterns. 
As illustrated in Fig.~\ref{signal}, the synthetic EMG patterns generated in the latent space closely resemble actual measured patterns, enabling the classifier to effectively learn discriminative features for combined motions.

The latent space visualization in Fig.~\ref{real_latent} provides crucial insights into why our method succeeds.
The structured embedding, where combined motions are positioned near the intermediate regions between their constituent basic motions, enables reliable interpolation for synthetic data generation.
This spatial organization results from the explicit regularization imposed by our mix-consistency loss $\mathcal{L}_{\text{mix}}$.
% This spatial organization results from the explicit regularization imposed by our mix-consistency loss $\mathcal{L}_{\text{mix}}$. 
The ablation study in Fig.\ref{accuracy_ablation} confirms the critical role of this loss component, as the baseline model without $\mathcal{L}_{\text{mix}}$ achieved only 40\% accuracy, similar to the Mixup method. 
This indicates that the VAE alone is insufficient to create the desired latent space structure, and the mix-consistency loss is essential for ensuring that learned representations respect the compositional nature of combined motions.
% Therefore, further improvements in the generative model, such as enhancing the reconstruction capability of the VAE and increasing robustness to noise, are expected to yield higher classification accuracy.

% Additionally, incorporating a conditional variational autoencoder, which reconstructs data conditioned on motion labels, may enable more stable and label-consistent data generation.
% This approach could potentially allow for accurate synthetic data generation without relying on explicit latent space structuring mechanisms such as the mix-consistency loss.

Despite these improvements, our method exhibits an approximately 10\% accuracy gap compared to the fully-supervised baseline (Table~\ref{table_accuracy}). 
This gap primarily stems from imperfect signal reconstruction, as evidenced by noise and amplitude variations in the synthetic patterns (Fig.~\ref{signal}), and the current requirement for some combined motion data during VAE training. 
Future work should focus on cross-subject transfer learning, where VAE models pre-trained on multiple subjects could enable synthetic data generation for new users with only basic motion data, thereby eliminating the need for individual combined motion collection and enhancing practical applicability.

% Another key challenge lies in the current experimental setup, where the VAE is trained using data from the same subject who performs the classification task. 
% Moreover, while the method aims to avoid direct collection of combined motion data, a portion of such data is still measured during the training phase.
% As a result, the data acquisition burden may not differ significantly from that of directly collecting combined motion data.

% Nonetheless, this study represents a crucial first step toward reducing user burden by demonstrating that high-quality synthetic data can be generated in structured latent space.
% This finding suggests that, once the latent space is properly structured, the generation of diverse, realistic combined motion signals becomes feasible.

% In future work, our aim is to extend this framework by enabling the transfer of the learned synthesis process across subjects.
% If the VAE can be trained or adapted using data from a limited number of source subjects, it may be possible to synthesize high-quality combined motion signals for new users with minimal calibration.
% This would eliminate the need to collect combined motion data from each individual, paving the way for more scalable and user-friendly EMG-based motion classification systems.

\section{Conclusion}
This study proposed a VAE-based method for recognizing unseen combined motions from EMG signals through synthetic data generation in a structured latent space. 
By introducing the mix-consistency loss, our approach ensures that combined motions are embedded between their constituent basic motions, enabling realistic synthetic pattern generation for classifier training.
% Our approach employs a VAE to learn a latent representation of EMG signals and introduces the mix-consistency loss to ensure that combined motions are embedded between their constituent basic motions. 
% This enables the generation of realistic synthetic EMG patterns for combined motions, which are then used to train a motion classifier alongside real basic motion data. 
% As a result, the model can recognize a wide range of motions from limited training data.

Experimental validation with eight participants demonstrated substantial improvements in combined motion classification accuracy, achieving 78.34\% compared to 44.46\% for conventional input-space synthesis methods.
The mix-consistency loss proved  essential for structuring the latent space, with ablation studies confirming its critical role in performance enhancement.

While this proof-of-concept demonstrates the feasibility of latent space synthesis, the current requirement for same-subject combined motion data during VAE training limits immediate burden reduction benefits. 
Future work should focus on cross-subject transfer learning to enable practical deployment with minimal per-user calibration.

% In this study, we used data from all motions, including combined motions, to train the VAE for the purpose of verifying the effectiveness of the structured latent space.
% In future work, we aim to extend our approach to cross-subject scenarios. 
% Specifically, we plan to pretrain the VAE on a small set of participants using signals from all motion types, and then apply it to new subjects by collecting only basic motion data to generate synthetic combined motion signals.
% This would enable the construction of classifier with minimal calibration effort.

% To further enhance the reconstruction capability of the VAE, we will explore techniques such as data augmentation and adversarial generative networks. 
% Moreover, improving the generalization ability of synthetic data generation to accommodate inter-individual variability in EMG signals is a key challenge. 
% Addressing these issues will move us closer to realizing a flexible EMG interface capable of accurately recognizing combined motions across different users.

% As a next step, we aim to extend our method to inter-subject scenarios by pretraining the VAE on a small dataset including both basic and combined motions, and applying it to new users with only basic motion data.
% Additionally, we will explore techniques such as data augmentation and adversarial training to improve reconstruction quality and robustness to inter-subject variability. 
% These improvements will support the development of a flexible EMG interface capable of accurately recognizing complex motions across diverse users with minimal calibration.

\addtolength{\textheight}{-12cm}   % This command serves to balance the column lengths

%%%%%%%%%%%%%%%%%%%%%%%%%%%%%%%%%%%%%%%%%%%%%%%%%%%%%%%%%%%%%%%%%%%%%%%%%%%%%%%%

\bibliographystyle{IEEEtran}
\bibliography{SII} 

\end{document}